\documentclass[aps,pra,showpacs,twocolumn]{revtex4-1}
\usepackage{amssymb}
\usepackage{amsmath}
\usepackage{graphicx}
\usepackage{epsfig}

\begin{document}

\title{Wide-range tunable Dirac-cone band structure in a chiral-time
symmetric non-Hermitian system}
\author{S. Lin and Z. Song}
\email{songtc@nankai.edu.cn}
\affiliation{School of Physics, Nankai University, Tianjin 300071, China}
\begin{abstract}
We establish a connection between an arbitrary Hermitian tight-binding model
with chiral ($\mathcal{C}$) symmetry and its non-Hermitian counterpart with
chiral-time ($\mathcal{CT}$) symmetry. We show that such a kind of
non-Hermitian Hamiltonian is pseudo-Hermitian. The eigenvalues and
eigenvectors of the non-Hermitian Hamiltonian can be easily obtained from
those of its parent Hermitian Hamiltonian. It provides a way to generate a
class of non-Hermitian models with a tunable full real band structure by
means of additional imaginary potentials. We also present an illustrative
example that could achieve a cone structure from the energy band of a
two-layer Hermitian square lattice model.
\end{abstract}

\maketitle


\section{Introduction}

Extra imaginary potentials induce many unusual features even in certain
simple or trivial systems, which include quantum phase transition occurred
in a finite system \cite%
{Znojil1,Znojil2,Bendix,LonghiPRL,LonghiPRB1,Jin1,Znojil3,LonghiPRB2,LonghiPRB3,Jin2,Joglekar1,Znojil4,Znojil5,Zhong,Drissi,Joglekar2,Scott1,Joglekar3,Scott2,Tony}%
, unidirectional propagation and anomalous transport \cite%
{LonghiPRL,Kulishov,LonghiOL,Lin,Regensburger,Eichelkraut,Feng,Peng,Chang},
invisible defects \cite{LonghiPRA2010,Della,ZXZ}, coherent absorption \cite%
{Sun} and self sustained emission \cite%
{MostafazadehPRL,LonghiSUS,ZXZSUS,Longhi2015,LXQ}, loss-induced revival of
lasing \cite{PengScience}, as well as laser-mode selection \cite%
{FengScience,Hodaei}. Most of these phenomena are related to the critical
behaviours near exceptional or spectral singularity points. It opens a way
for exploring novel quantum states. The basis of such approaches is to seek
various non-Hermitian systems with exact solutions. Recently, the
graphene-like materials with Dirac cones at the Fermi energy and a number of
unique mechanical, electrical, and optical properties, have attracted much
attention \cite{RMP}. Its linear-Dirac dispersion makes it an active topic
in various research fields. However, for materials in nature, it is very hard
to realize experimentally with tuneable parameters. An artificial system,
such as photonic simulator, would provide a platform to simulate some
aspects in various band structures. Previous efforts mainly focus on the
Hermitian systems. A natural question would emerge that whether one can find
some artificial materials which have a cone band structure.

In this paper, we consider a method of constructing a variety of
non-Hermitian systems which have full real spectra. We focus on the
connection between an arbitrary Hermitian tight-binding model with chiral ($%
\mathcal{C}$) symmetry and its non-Hermitian counterpart with chiral-time ($%
\mathcal{CT}$) symmetry. We show that such a kind of non-Hermitian
Hamiltonian is pseudo-Hermitian. The obtained result indicates that the
eigenvalues and eigenvectors of the non-Hermitian Hamiltonian can be easily
obtained from those of its parent Hermitian Hamiltonian and the reality of
the spectrum is robust to the disorder. It also provides a way to generate a
class of non-Hermitian models with a tunable full real band structure by
means of additional imaginary potentials. We present an illustrative
example, which is a two-layer square lattice model. By adding staggered
imaginary potentials, exact result shows that a cone band structure can be
achieved.

The remainder of this paper is organized as follows. In Sec. \ref{Model and
formalism}, we present a general formalism\ for the solution of an arbitrary
non-Hermitian $\mathcal{CT}$-symmetric system. Sec. \ref{Valley structure}
is devoted to present an illustrative example of a two-layer square lattice
model. Finally, we present a summary and discussion in Sec. \ref{Summary}.

\section{Model and formalism}

\label{Model and formalism}The main interest of this work is focused on the
relation between an arbitrary Hermitian tight-binding model with $\mathcal{C}
$ symmetry and a non-Hermitian model which is constructed based on the
former by adding additional imaginary potentials. The latter is a
non-Hermitian counterpart of the former in the context of this work.

Consider the Hamiltonian of a non-Hermitian tight-binding model
\begin{equation}
H=H_{0}+H_{\gamma }  \label{H}
\end{equation}%
with%
\begin{eqnarray}
H_{0} &=&\sum_{i,j}J_{ij}\left\vert i\right\rangle _{\mathrm{A}}\left\langle
j\right\vert _{\mathrm{B}}+\mathrm{H.c.}, \\
H_{\gamma } &=&i\gamma \sum_{j}(\left\vert j\right\rangle _{\mathrm{A}%
}\left\langle j\right\vert _{\mathrm{A}}-\left\vert j\right\rangle _{\mathrm{%
B}}\left\langle j\right\vert _{\mathrm{B}}),
\end{eqnarray}%
on a bipartite lattice $\Lambda =2N$ which can be decomposed into two
sublattices $\Lambda _{\mathrm{A}}$ and $\Lambda _{\mathrm{B}}$. Here we
only consider the case with identical sublattice numbers $\Lambda _{\mathrm{A%
}}=\Lambda _{\mathrm{B}}=N$\ for simplicity. A schematic illustration of the
model is presented in Fig. 1(a).
\begin{figure}[tbp]
\includegraphics[ bb=30 530 490 800, width=0.49\textwidth, clip]{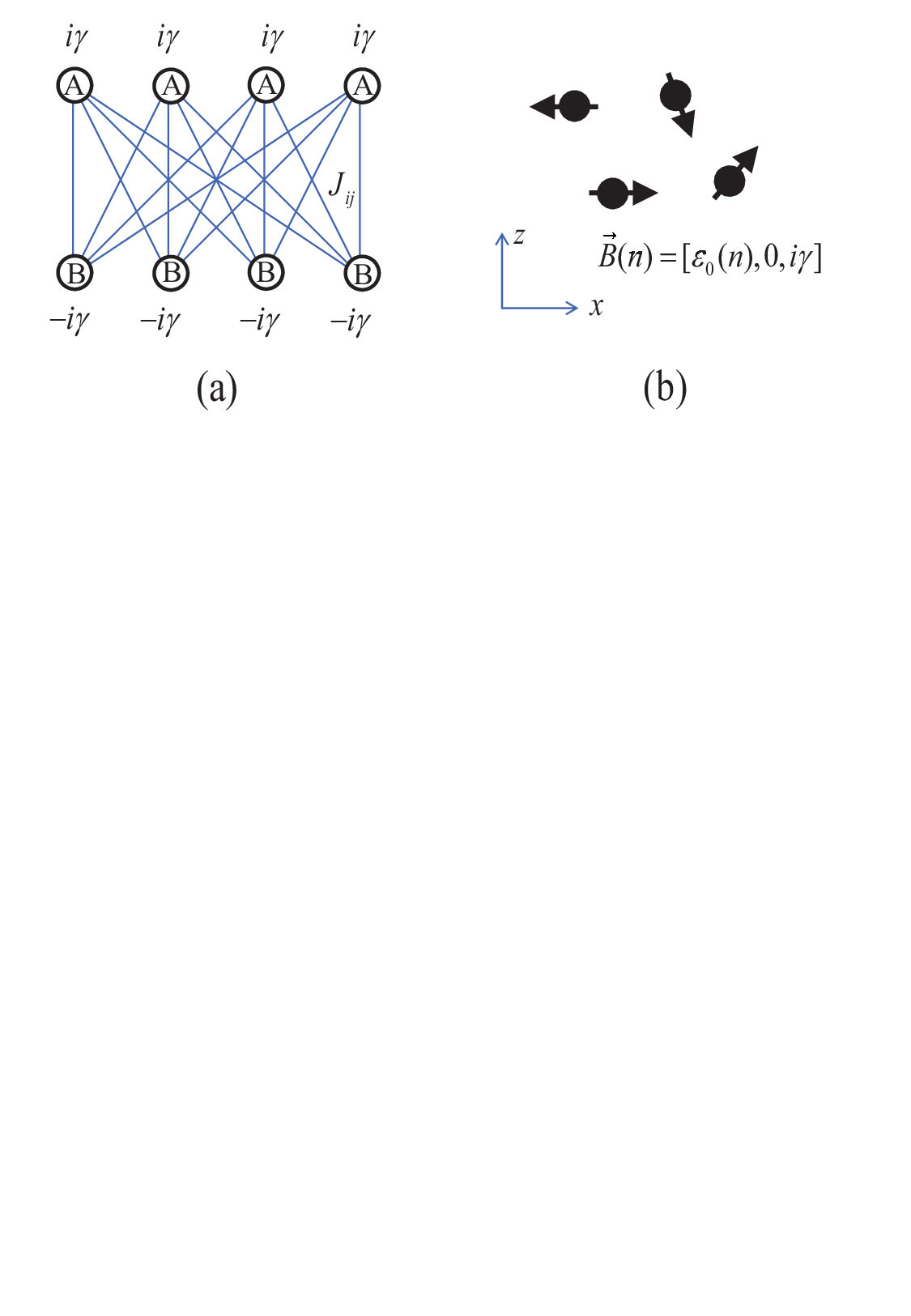}
\caption{ (color online). Schematics for the system with $\Lambda _{\mathrm{A%
}}=\Lambda _{\mathrm{B}}=4$ to illustrate the connection between the systems
of Eqs. (\protect\ref{H}) and (\protect\ref{spin-field}). (a) A bipartite
lattice consists of sublattices \textrm{A} and \textrm{B}, which are
connected by bond $J_{ij}$ which is across the $i$th site in sublattice
\textrm{A} and the $j$th site in sublattice \textrm{B}. In the absence of
imaginary potentials, i.e., $\protect\gamma =0$, it has $\mathcal{C}$
symmetry, which ensures that the system has the spectrum $\pm \protect%
\varepsilon _{0}(n)$ ($n=1,2,3,4$). In the presence of imaginary potentials,
it has $\mathcal{CT}$ symmetry and becomes a pseudo-Hermitian system. (b) An
ensemble of non-interacting half spins in a complex external magnetic field.
It is an equivalent system of (a) when the local magnetic field for spin $n$
has the form $\protect\overrightarrow{B}(n)$.}
\label{fig1}
\end{figure}
The Hamiltonian $H_{0}$\ has both $\mathcal{C}$ and time-reversal ($\mathcal{%
T}$)\ symmetries, i.e.,%
\begin{equation}
\mathcal{C}H_{0}\mathcal{C}^{-1}=-H_{0},\mathcal{T}H_{0}\mathcal{T}%
^{-1}=H_{0},
\end{equation}%
where the operators $\mathcal{C}$\ and $\mathcal{T}$\ are defined as
\begin{eqnarray}
&&\mathcal{C}\left\vert j\right\rangle _{\mathrm{A}}=\left\vert
j\right\rangle _{\mathrm{A}},\text{ }\mathcal{C}\left\vert j\right\rangle _{%
\mathrm{B}}=-\left\vert j\right\rangle _{\mathrm{B}}, \\
&&\mathcal{T}\sqrt{-1}\mathcal{T}^{-1}=-\sqrt{-1}.
\end{eqnarray}%
The Hamiltonian $H$\ has $\mathcal{CT}$\ symmetry, i.e.,
\begin{eqnarray}
&&\mathcal{C}H\mathcal{C}^{-1}\neq -H,\mathcal{T}H\mathcal{T}^{-1}\neq H, \\
&&\mathcal{CT}H\mathcal{T}^{-1}\mathcal{C}^{-1}=-H.
\end{eqnarray}%
We note that Hamiltonian $H_{0}$ has $\mathcal{C}$ symmetry, which is broken
in its non-Hermitian counterpart $H$ in the presence of imaginary potentials
$H_{\gamma }$. The situation here is a little different from the case
associated with parity-time ($\mathcal{PT}$) symmetry, where the combined
operator $\mathcal{PT}$ commutes with the Hamiltonian. In quantum mechanics,
we say that a Hamiltonian $H$\ has a symmetry represented by a operator $%
\mathcal{U}$\ if $[H,\mathcal{U}]=0$. The word \textquotedblleft
symmetry\textquotedblright\ is also used in a different sense in condensed
matter physics. We say that a system with Hamiltonian $H$\ has chiral
symmetry, if $\{H,\mathcal{C}\}=0$. The physics of $\mathcal{C}$\ depends on
the model discussed \cite%
{Asboth2016lnp,Malzard2015prl,Guo2015prb,ChenS2015prb,Peng2016np,Lee2016prl}%
. Here we emphasize \textquotedblleft chiral symmetry\textquotedblright\ due
to its anticommutation relation with its Hamiltonians.\textbf{\ }%
Specifically, the anticommutation relation between operators $\mathcal{CT}$
and $H$ results in the equations%
\begin{eqnarray}
H\left\vert \psi \right\rangle &=&\varepsilon \left\vert \psi \right\rangle ,
\label{CT_definition1} \\
H\mathcal{CT}\left\vert \psi \right\rangle &=&-\varepsilon ^{\ast }\mathcal{%
CT}\left\vert \psi \right\rangle .  \label{CT_definition2}
\end{eqnarray}
However, the $\mathcal{CT}$ symmetry is like anti $\mathcal{PT}$ symmetry
\cite{Peng2016np}. Actually, an anti-$\mathcal{PT}$-symmetric Hamiltonian
can be simply constructed from a conventional $\mathcal{PT}$-symmetric
Hamiltonian by multiplying $i$. Here we present two tables demonstrate the
difference and connection between $\mathcal{CT}$\ and $\mathcal{PT}$
symmetry.

\begin{table}[tbp]
\caption{The difference and connection between $\mathcal{CT}$ and $\mathcal{%
PT}$\ symmetry.}
\label{Table 1(a)}
\begin{center}
\renewcommand\arraystretch{1.5}
\par
\begin{tabular}{ccc}
\hline\hline
\parbox{1.9cm}{$H\left\vert \psi \right\rangle =\varepsilon \left\vert \psi
\right\rangle$} & \parbox{3.3cm} {$\mathcal{PT}$} &
\parbox{3.2cm}
{$\mathcal{CT}$} \\ \hline
\parbox{1.9cm}{$\text{symmetry}$} &
\parbox{3.3cm} {$\left[
\mathcal{PT},H\right] =0$} &
\parbox{3.2cm} {$\left\{ \mathcal{CT},H\right\}
=0$} \\
\parbox{1.9cm}{$\text{real }\varepsilon$} &
\parbox{3.3cm}
{$H\mathcal{PT}\left\vert \psi \right\rangle=\varepsilon
\mathcal{PT}\left\vert \psi \right\rangle$} &
\parbox{3.2cm}
{$H\mathcal{CT}\left\vert \psi \right\rangle =-\varepsilon
\mathcal{CT}\left\vert \psi\right\rangle$} \\
\parbox{1.9cm}{$\text{imaginary }\varepsilon$} &
\parbox{3.3cm}
{$H\mathcal{PT}\left\vert \psi \right\rangle=-\varepsilon
\mathcal{PT}\left\vert \psi \right\rangle$} &
\parbox{3.2cm}
{$H\mathcal{CT}\left\vert \psi \right\rangle =\varepsilon
\mathcal{CT}\left\vert \psi\right\rangle$} \\ \hline\hline
\parbox{1.5cm}{} & \parbox{2cm}{(a)} & \parbox{1.5cm}{}%
\end{tabular}%
\end{center}
\end{table}

\begin{table}[tbp]
\label{Table 1(b)}
\par
\begin{center}
\renewcommand\arraystretch{1.5} \renewcommand{\thetable}{\CJKnumber{%
\value{table}}}
\par
\begin{tabular}{ccc}
\hline\hline
\parbox{1.9cm}{$H^{^{\prime }}=iH$} & \parbox{3.3cm} {$\mathcal{PT}$} & %
\parbox{3.2cm} {$\mathcal{CT}$} \\ \hline
\parbox{1.9cm}{$\text{symmetry}$} &
\parbox{3.3cm} {$\left\{
\mathcal{PT},H^{^{\prime }}\right\} =0$} &
\parbox{3.2cm}
{$\left[\mathcal{CT},H^{^{\prime }}\right] =0$} \\
\parbox{1.9cm}{$\text{real }\varepsilon$} &
\parbox{3.3cm} {$H^{^{\prime }}\mathcal{PT}\left\vert \psi
\right\rangle =-\varepsilon \mathcal{PT}\left\vert \psi \right\rangle$} &
\parbox{3.2cm} {$H^{^{\prime }}\mathcal{CT}\left\vert \psi \right\rangle
=\varepsilon\mathcal{CT}\left\vert \psi \right\rangle$} \\
\parbox{1.9cm}{$\text{imaginary }\varepsilon$} &
\parbox{3.3cm} {$H^{^{\prime }}\mathcal{PT}\left\vert \psi
\right\rangle =\varepsilon \mathcal{PT}\left\vert \psi \right\rangle$} &
\parbox{3.2cm} {$H^{^{\prime }}\mathcal{CT}\left\vert \psi \right\rangle
=-\varepsilon\mathcal{CT}\left\vert \psi \right\rangle$} \\ \hline\hline
\parbox{1.5cm}{} & \parbox{2cm}{(b)} & \parbox{1.5cm}{}%
\end{tabular}%
\end{center}
\end{table}
Since the relation $\{\mathcal{CT},H\}=0$\ cannot guarantee
operators $\mathcal{CT}$\ and $H$\ possess a common complete set of
eigensates, it is difficult to define the $\mathcal{CT}$\ symmetry of a
state $\left\vert \psi \right\rangle $. In order to define the $\mathcal{CT}$%
\ symmetry of a state, we consider the operator $iH$\ which obeys the
relation $[\mathcal{CT},iH]=0$. Then $\mathcal{CT}$\ and $H$\ can have a
common complete set of eigensates. The $\mathcal{CT}$\ symmetry of a state $%
\left\vert \psi \right\rangle $\ is defined as usual, $\mathcal{CT}%
\left\vert \psi \right\rangle =c\left\vert \psi \right\rangle $.
Accordingly, in the exact $\mathcal{CT}$-symmetric region, all the
eigenstate obeys $\mathcal{CT}\left\vert \psi \right\rangle =c\left\vert
\psi \right\rangle $\ and $iH$\ has fully real spectrum. For the concerned
model, the eigenenergy of $H$\ is either real or pure imaginary. When all
the eigenstates break the $\mathcal{CT}$\ symmetry, i.e., $\mathcal{CT}%
\left\vert \psi \right\rangle \neq c\left\vert \psi \right\rangle $, the
Hamiltonian has fully real spectrum, and $\left\vert \psi \right\rangle $\
and $\mathcal{CT}\left\vert \psi \right\rangle $\ have the opposite real
eigenenergies.

Now we investigate the Hamiltonian $H$ in a pseudo spin representation. We
will show that $H$ is a pseudo-Hermitian Hamiltonian and there is a simple
relation between the spectra of $H$ and $H_{0}$. Due to the $\mathcal{C}$
symmetry, the Hamiltonian $H_{0}$\ can be diagonalized as the form%
\begin{equation}
H_{0}=\sum_{n=1}^{N}\varepsilon _{0}(n)(\left\vert \varphi
_{n}^{+}\right\rangle \left\langle \varphi _{n}^{+}\right\vert -\left\vert
\varphi _{n}^{-}\right\rangle \left\langle \varphi _{n}^{-}\right\vert ),
\end{equation}%
where $\varepsilon _{0}(n)>0$ is the positive energy spectrum with $n\in
\lbrack 1,N]$, and%
\begin{equation}
\left\vert \varphi _{n}^{\pm }\right\rangle =\frac{1}{\sqrt{2}}(\left\vert
\phi _{n}\right\rangle _{\mathrm{A}}\pm \left\vert \phi _{n}\right\rangle _{%
\mathrm{B}}),
\end{equation}%
are eigenstates with eigenenergies $\pm \varepsilon _{0}(n)$. Here states $%
\left\vert \phi _{n}\right\rangle _{\mathrm{A}}$\ and $\left\vert \phi
_{n}\right\rangle _{\mathrm{B}}\ $are single-particle states with particle
probability only distributed on sublattices A and B, respectively. Due to
the $\mathcal{C}$ symmetry of $H_{0}$, it is easy to check that $\left\vert
\varphi _{n}^{-}\right\rangle =\mathcal{C}\left\vert \varphi
_{n}^{+}\right\rangle $. One can express the Hamiltonian in the
representation of pseudo spins%
\begin{equation}
H_{0}=\sum_{n=1}^{N}\varepsilon _{0}(n)\sigma _{n}^{x},
\end{equation}%
where
\begin{equation}
\sigma _{n}^{x}=\left\vert \phi _{n}\right\rangle _{\mathrm{B}}\left\langle
\phi _{n}\right\vert _{\mathrm{A}}+\left\vert \phi _{n}\right\rangle _{%
\mathrm{A}}\left\langle \phi _{n}\right\vert _{\mathrm{B}}
\end{equation}%
is the $x$-component of the Pauli matrix. Accordingly, we could rewrite the
Hamiltonian $H$\ as the form%
\begin{equation}
H=\sum_{n=1}^{N}\overrightarrow{B}(n)\cdot \overrightarrow{\sigma }_{n},
\label{spin-field}
\end{equation}%
which describes an ensemble of non-interacting half spins in a complex
external magnetic field. Here the field and the Pauli matrices are%
\begin{eqnarray}
\overrightarrow{B}(n) &=&[\varepsilon _{0}(n),0,i\gamma ], \\
\sigma _{n}^{z} &=&\left\vert \phi _{n}\right\rangle _{\mathrm{A}%
}\left\langle \phi _{n}\right\vert _{\mathrm{A}}-\left\vert \phi
_{n}\right\rangle _{\mathrm{B}}\left\langle \phi _{n}\right\vert _{\mathrm{B}%
}, \\
\sigma _{n}^{y} &=&i\left\vert \phi _{n}\right\rangle _{\mathrm{B}%
}\left\langle \phi _{n}\right\vert _{\mathrm{A}}-i\left\vert \phi
_{n}\right\rangle _{\mathrm{A}}\left\langle \phi _{n}\right\vert _{\mathrm{B}%
}.
\end{eqnarray}%
Based on this analysis, the eigenstates and eigenenergies of Hamiltonian $H$%
\ are%
\begin{eqnarray}
\left\vert \psi _{n}^{\pm }\right\rangle &=&\frac{1}{\sqrt{\Omega _{\pm }}}%
(\left\vert \phi _{n}\right\rangle _{\mathrm{A}}\pm e^{\mp i\theta
}\left\vert \phi _{n}\right\rangle _{\mathrm{B}}), \\
\varepsilon (n) &=&\pm \sqrt{\lbrack \varepsilon _{0}(n)]^{2}-\gamma ^{2}},
\label{En}
\end{eqnarray}%
where $\theta =\arccos \sqrt{1-\left[ \gamma /\varepsilon _{0}(n)\right] ^{2}%
}$ and the Dirac normalized coefficients are $\Omega _{\pm }=1+\exp \left(
\pm 2\mathrm{Im}\theta \right) $.

This result has many implications. (i) Non-Hermitian Hamiltonian $H$\ is
pseudo-Hermitian, since it has either a real spectrum or else its complex
eigenvalues always occur in complex conjugate pairs \cite{MostafazadehJPA}.
(ii) It explicitly connects the complete set $\{\varepsilon (n),$ $%
\left\vert \psi _{n}^{\pm }\right\rangle \}$\ to $\{\pm \varepsilon _{0}(n),$
$\left\vert \varphi _{n}^{\pm }\right\rangle \}$. Only an extra phase is
added in $\left\vert \psi _{n}^{\pm }\right\rangle $\ from $\left\vert
\varphi _{n}^{\pm }\right\rangle $, which indicates that the two states have
the same Dirac\ probability distribution when $\varepsilon (n)$\ is real.\
(iii) The exceptional points occur at $\gamma =\gamma _{\mathrm{c}}=\pm
\varepsilon _{0}(n)$,\ which correspond to the $\mathcal{CT}$ symmetry
breaking of states $\left\vert \psi _{n}^{\pm }\right\rangle $. It allows a
variety of non-Hermitian models with a wide range of disorder parameters to
have a full real spectrum and the modulation of band structure is due to the
non-Hermiticity. In the next section, we will show its\ application in an
example.

To demonstrate these features, we consider an example model, a generalized
non-Hermitian Rice-Mele model, which has been investigated in Ref. \cite%
{HWHPRA,LSPRA}. The corresponding Hermitian Hamiltonian has the form
\begin{equation}
H_{0}=\sum_{j=1}^{N}(J_{2j-1}\left\vert j\right\rangle _{\mathrm{A}%
}\left\langle j\right\vert _{\mathrm{B}}+J_{2j}\left\vert j\right\rangle _{%
\mathrm{B}}\left\langle j+1\right\vert _{\mathrm{A}})+\mathrm{H.c.}
\end{equation}%
with the periodic boundary condition $\left\vert 2N+1\right\rangle _{\mathrm{%
A}}=\left\vert 1\right\rangle _{\mathrm{A}}$. The hopping amplitude between
two sublattices is $J_{j}=1+\left( -1\right) ^{j}\delta ,$ where $\delta $
is the dimerization factor. The generalized non-Hermitian Rice-Mele
Hamiltonian has been completely solved and the obtained result can be
recovered by the present method. In this case, we have%
\begin{eqnarray}
\varepsilon (k) &=&\pm \sqrt{\lbrack \varepsilon _{0}(k)]^{2}-\gamma ^{2}},
\\
\varepsilon _{0}(k) &=&2\sqrt{\delta ^{2}+\left( 1-\delta ^{2}\right) \cos
^{2}\left( k/2\right) },
\end{eqnarray}%
with $k=2\pi n/N$, $n\in \lbrack 1,N]$. In the absence of $\gamma $, the
energy gap is $4\delta $, which determines the exceptional point occurring
at $\gamma =\gamma _{\mathrm{c}}=\pm 2\delta $\ for the non-Hermitian
Rice-Mele Hamiltonian. In other words, the energy gap of $H_{0}$ protects
the $\mathcal{CT}$ symmetry\ of the eigenstates of $H$. This is still true
in the presence of noise in $J_{j}$. In contrast to a $\mathcal{PT}$
symmetric system, the reality of the spectrum for a $\mathcal{CT}$ symmetric
one is more robust to the disorder of coupling constants.

\begin{figure}[tbp]
\includegraphics[ bb=44 566 440 806, width=0.40\textwidth, clip]{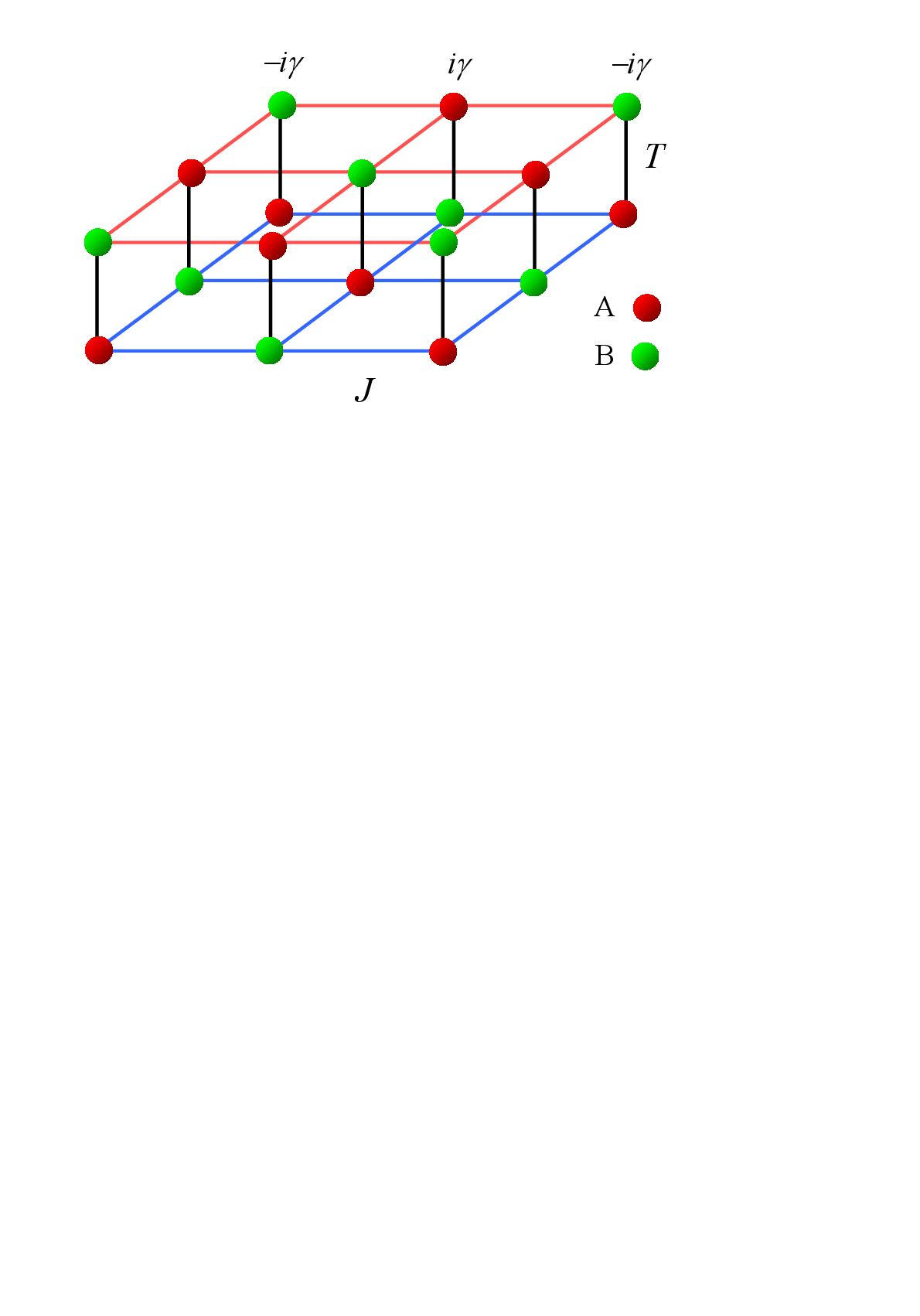}
\caption{(color online). (a) Schematic illustration of a bilayer square
lattice with staggered imaginary potentials. The two sublattices are denoted
by A (red) and B (green), respectively. The intra and interlayer hopping
strengths are $J$ and $T$, respectively. For $\protect\gamma =0$, the system
has both $\mathcal{C}$ and $\mathcal{T}$ symmetries, while nonzero $\protect%
\gamma $\ breaks the $\mathcal{C}$ symmetry but maintains the $\mathcal{CT}$
symmetry. The additional staggered imaginary potentials make the simple
lattice have a tunable cone band structure.}
\label{fig2}
\end{figure}
\begin{figure*}[tbp]
\includegraphics[ bb=67 20 740 576, width=0.329\textwidth, clip]{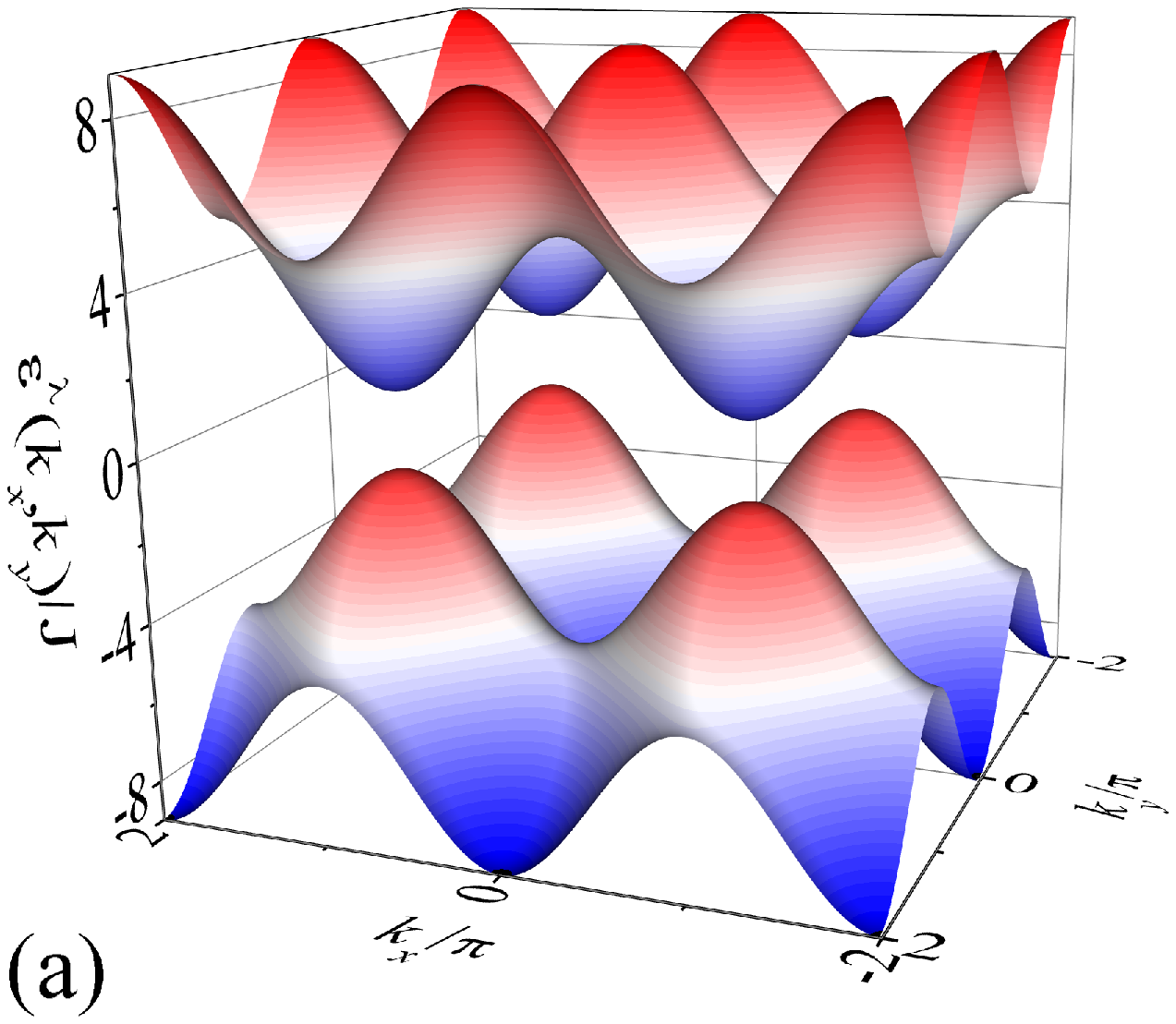} %
\includegraphics[ bb=67 20 740 576, width=0.329\textwidth, clip]{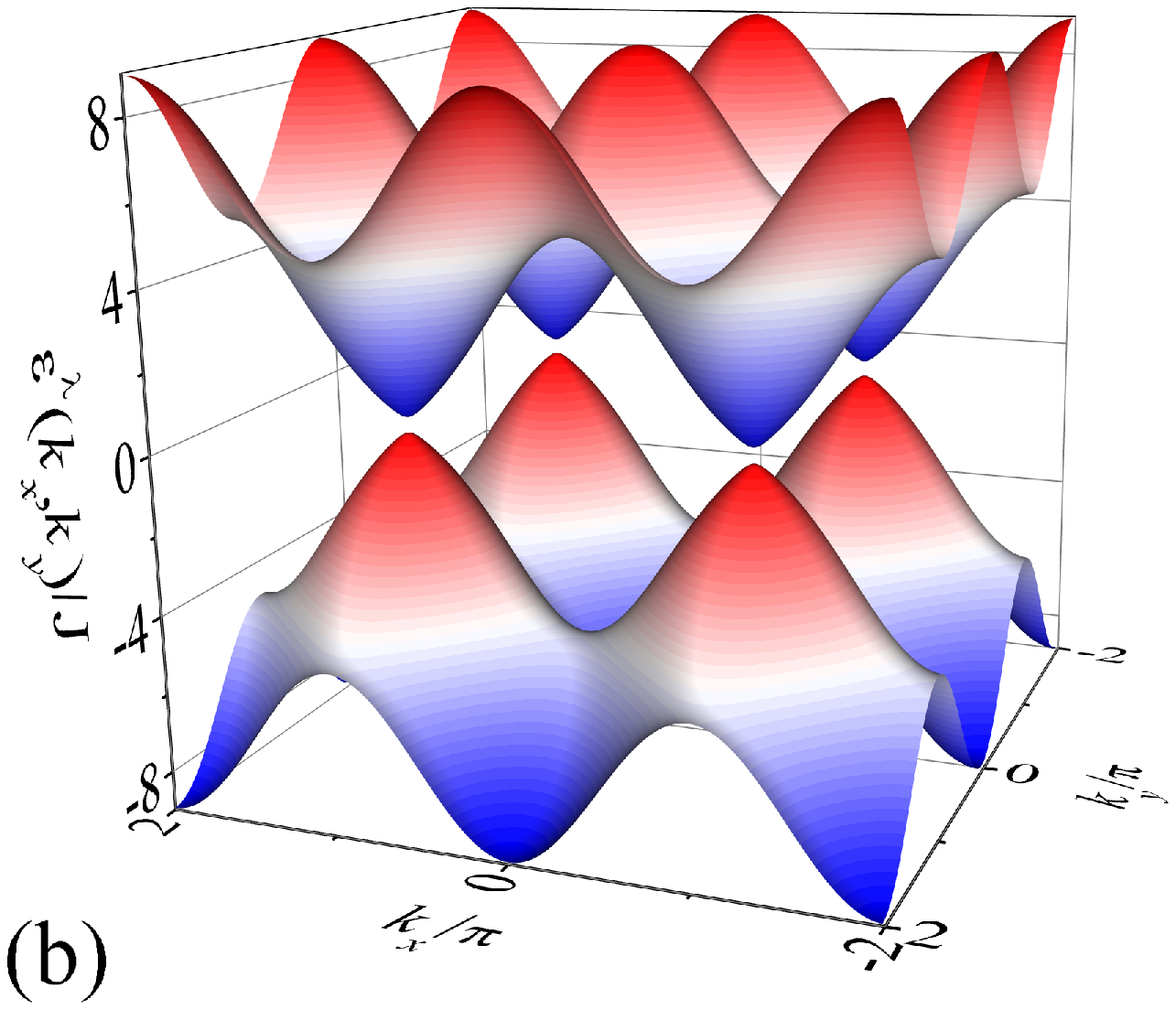} %
\includegraphics[ bb=67 20 740 576, width=0.329\textwidth, clip]{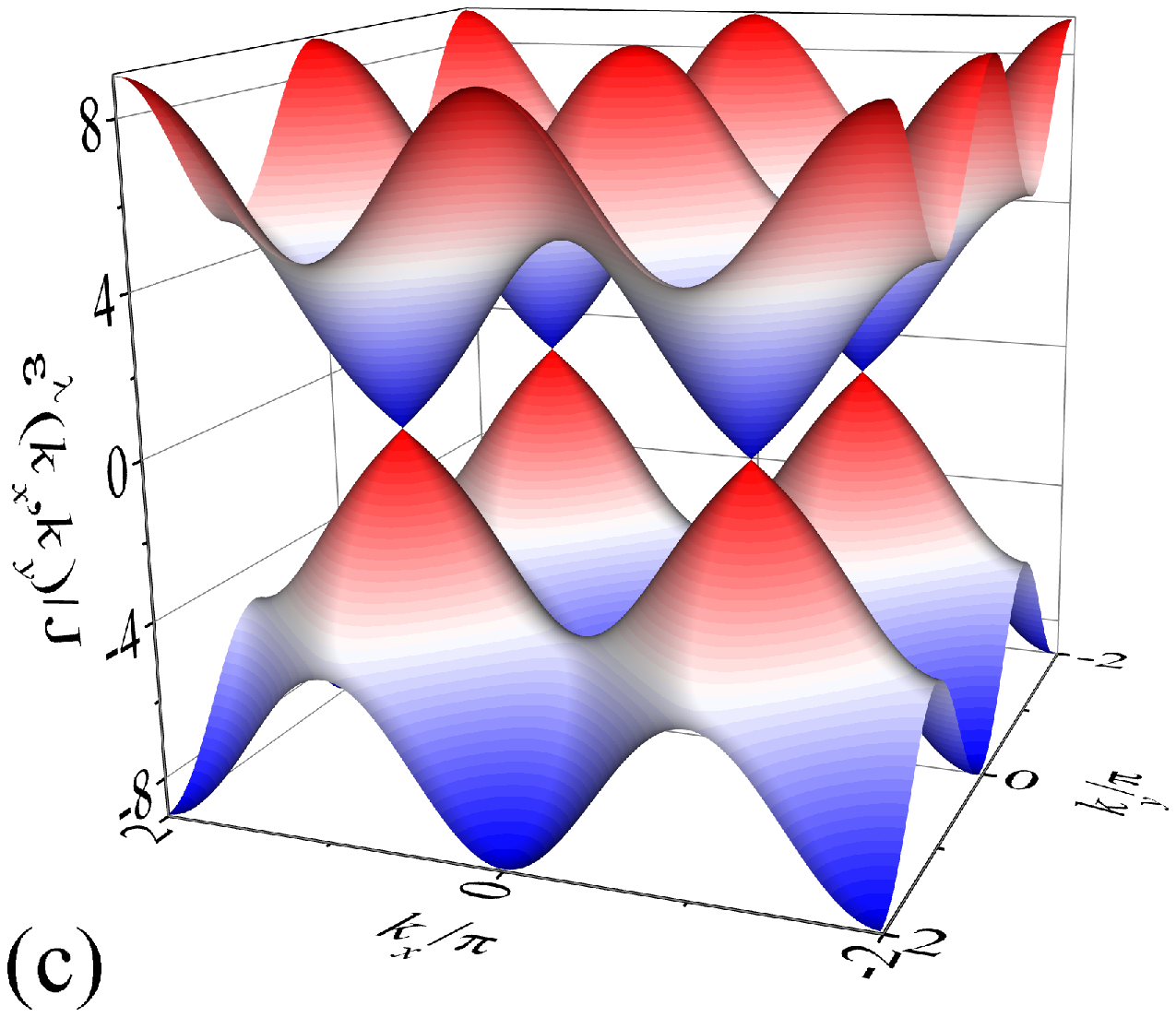}
\caption{(color online). 3D plots of band structures of bilayer square
lattice with periodic boundary condition. Staggered imaginary potentials $%
\pm i\protect\gamma $ are applied throughout the lattice and here we set $%
\protect\lambda =\pm $. The parameters are (a) $T=5J,\protect\gamma =0,$ (b)
$T=5J,\protect\gamma =0.98J,$ (c) $T=5J,\protect\gamma =J.$ Here the key
difference between the cases for (a) and (c) is that the dispersion relation
in the bottom of band is quadratic for (a) but linear (cone) for (c). The
band structure in case (a) is trivial in the context that Dirac cone in
lattice system has been shown to exhibit some novel features. Although a
Hermitian system can support Dirac dispersion (e.g., honeycomb lattice), the
speed of electron (slope of the cone) is not tunable.}
\label{fig3}
\end{figure*}

\section{Cone structure}

\label{Valley structure}

The connection between $H_{0}$ and $H$ can be employed to modulate the band
structure of $H$, which has some intriguing properties induced by the
non-Hermitian term $H_{\gamma }$. In traditional condensed matter theories,
the energy band structure plays a crucial role in the theory of electron
conductivity in the solid state and explains why materials can be classified
as insulators, conductors and semiconductors. Moreover, much attention has
been paid to the honeycomb lattice \cite{RMP}, which is relevant to high
electron mobility and topological phase, as exemplified by the graphene.

In the Hermitian realm, the band structures of most kinds of systems have
been\ well studied. Nevertheless, non-Hermitian parameters may induce an
unusual band structure which is difficult to achieve in a Hermitian system.
As an example, Eq. (\ref{En}) provides a way to accomplish this task that
imaginary potentials can deform the shape of a given band structure without
altering its topology except the situation when the system contains the
exceptional points. In the following, we will present an example which
realizes a cone structure on a square lattice.

We consider a bilayer square lattice model which is shown in Fig. \ref{fig2}%
. The corresponding Hermitian Hamiltonian has the form
\begin{eqnarray}
H_{0} &=&H_{1}+H_{2}+H_{12}, \\
H_{\lambda } &=&J\sum_{j,l=1}^{N}\left\vert \lambda ,j,l\right\rangle
(\left\langle \lambda ,j+1,l\right\vert  \notag \\
&&+\left\langle \lambda ,j,l+1\right\vert )+\mathrm{H.c.}, \\
H_{12} &=&T\sum_{j,l=1}^{N}\left\vert 1,j,l\right\rangle \left\langle
2,j,l\right\vert +\mathrm{H.c.},  \label{bilayer}
\end{eqnarray}%
where $\lambda =1$ or $2$ is the index that respectively labels the position
in the top or bottom layers, and $(j,l)$ is the in-plane site index.
Parameters $J$ and $T$ of this model are intra and interlayer hopping
strengths. In this paper, we only consider the case of $T>4J$. And the
distribution of imaginary potentials is given as the form%
\begin{equation}
H_{\gamma }=i\gamma \sum_{\lambda =1}^{2}\sum_{j,l=1}^{N}(-1)^{\lambda
+j+l}\left\vert \lambda ,j,l\right\rangle \left\langle \lambda
,j,l\right\vert .
\end{equation}%
The Hamiltonian $H_{0}$\ can be easily diagonalized via Fourier
transformation. Let us consider an individual rung, i.e. two sites with the
same in-plane site index on the opposite layers. An occupied rung has two
possible states that are bond and antibond states. The bond (antibond) state
of a rung can only be transited to the bond (antibond) state next to it.
Therefore it can be decomposed into two independent single layer square
lattices with on-site potentials $T$\ and $-T$, respectively. The spectra
and eigenvectors are%
\begin{equation}
\varepsilon _{0}^{\pm }(k_{x},k_{y})=\pm \left\{ 2J[\cos (k_{x})+\cos
(k_{y})]+T\right\} ,
\end{equation}%
and%
\begin{equation}
\left\vert \varphi ^{\pm }(k_{x},k_{y})\right\rangle =\sum_{j,l=1}^{N}\frac{%
e^{i\left( k_{x}j+k_{y}l\right) }}{N\sqrt{2}}\left( \left\vert \lambda _{%
\mathrm{A}},j,l\right\rangle \pm \left\vert \lambda _{\mathrm{B}%
},j,l\right\rangle \right) ,
\end{equation}%
where $\pm $\ denotes the two independent single layers, and $%
k_{x}=2n_{x}\pi /N$, $k_{y}=2n_{y}\pi /N$ with $n_{x}$, $n_{y}\in \left[ 1,N%
\right] $. States\textbf{\ }$\left\vert \lambda _{\mathrm{A}%
},j,l\right\rangle $ and $\left\vert \lambda _{\mathrm{B}},j,l\right\rangle $%
\textbf{\ }are the position states of sublattices A and B with the layer
labels $\lambda _{\mathrm{A}}=[3+\left( -1\right) ^{j+l}]/2$ and $\lambda _{%
\mathrm{B}}=[3-\left( -1\right) ^{j+l}]/2$. This band structure is trivial,
but it would be a good parent to construct a cone structure by adding
staggered imaginary potentials. Now we consider the corresponding
non-Hermitian Hamiltonian $H=H_{0}+H_{\gamma }$. According to the above
result, the spectra and eigenvectors of $H$\ are
\begin{equation}
\varepsilon ^{\pm }(k_{x},k_{y})=\pm \sqrt{\lbrack \varepsilon _{0}^{\pm
}(k_{x},k_{y})]^{2}-\gamma ^{2}},  \label{E2D}
\end{equation}%
and%
\begin{equation}
\left\vert \psi ^{\pm }(k_{x},k_{y})\right\rangle =\sum_{j,l=1}^{N}\frac{%
e^{i\left( k_{x}j+k_{y}l\right) }}{N\sqrt{\Omega _{\pm }}}\left( \left\vert
\lambda _{\mathrm{A}},j,l\right\rangle \pm e^{\mp i\theta }\left\vert
\lambda _{\mathrm{B}},j,l\right\rangle \right) ,
\end{equation}%
where
\begin{equation}
\theta =\arccos \sqrt{1-(\gamma /\varepsilon _{0}^{\pm })^{2}}
\end{equation}%
is real when the symmetry is not broken. In the exact $\mathcal{CT}$%
-symmetric region, there are local maxima (minima) on the valence
(conduction) band at points $(k_{x}^{\mathrm{c}},k_{y}^{\mathrm{c}})=(\sigma
\pi ,\sigma ^{\prime }\pi )$ ($\sigma ,\sigma ^{\prime }=$ odd). The energy
band gap is $2\sqrt{(T-4J)^{2}-\gamma ^{2}}$ and the exceptional points
occur at $\gamma =\gamma _{\mathrm{c}}=\varepsilon _{0}(k_{x}^{\mathrm{c}%
},k_{y}^{\mathrm{c}})$ $=T-4J$. In the vicinity of $k_{\mathrm{c}}$ and
considering the case $0<\gamma _{\mathrm{c}}-\gamma \ll J$, we have an
approximate relation
\begin{equation}
\frac{\left( k_{x}-k_{x}^{\mathrm{c}}\right) ^{2}}{a^{2}}+\frac{\left(
k_{y}-k_{y}^{\mathrm{c}}\right) ^{2}}{b^{2}}-\frac{\left( \varepsilon ^{\pm
}\right) ^{2}}{c^{2}}=-1,  \label{hyperboloid}
\end{equation}%
with $c=\sqrt{\gamma _{\mathrm{c}}^{2}-\gamma ^{2}}$ and $a=b=c/\sqrt{%
2J\gamma _{\mathrm{c}}}$, which indicates that the band structure is a
hyperboloid of two sheets. For $\gamma =\gamma _{\mathrm{c}}$, it reduces to
a Dirac cone. Note that the difference between the cases for $\gamma =0$\
and $\gamma =\gamma _{\mathrm{c}}$\ is that the dispersion relation in the
bottom of band is quadratic for $\gamma =0$\ but linear (cone) for $\gamma
=\gamma _{\mathrm{c}}$. Although a Hermitian system can support Dirac
dispersion (e.g., honeycomb lattice), the speed of electron (slope of the
cone) is less tunable. In contrast, the group velocity at the
linear region for our model is
\begin{equation}
v_{\mathrm{g}}=J\sqrt{2\left( T/J-4\right) },
\end{equation}%
which indicates that $v_{\mathrm{g}}$\ strongly depends on the ratio of $J$
and $T$ ($\gamma _{\mathrm{c}}=T-4J$), while it only depends on the hopping
strength in a honeycomb lattice. In this sense, imaginary extension may make
something easier to achieve than that in a Hermitian system. Furthermore, it
seems that it has a similar band structure with that of graphene near the
zero-energy plane. The difference between them is that the vertices of the
cone of graphene are degenerate points, while the ones in the present model
are exceptional points. For $\gamma <\gamma _{\mathrm{c}}$, the energy gap
and the group velocity are tunable by $\gamma $, $J$, and $T$. The cone band
structures for different $\gamma $\ are plotted in Fig. \ref{fig3}.
\begin{figure}[tbp]
\includegraphics[ bb=8 11 542 223, width=0.48\textwidth, clip]{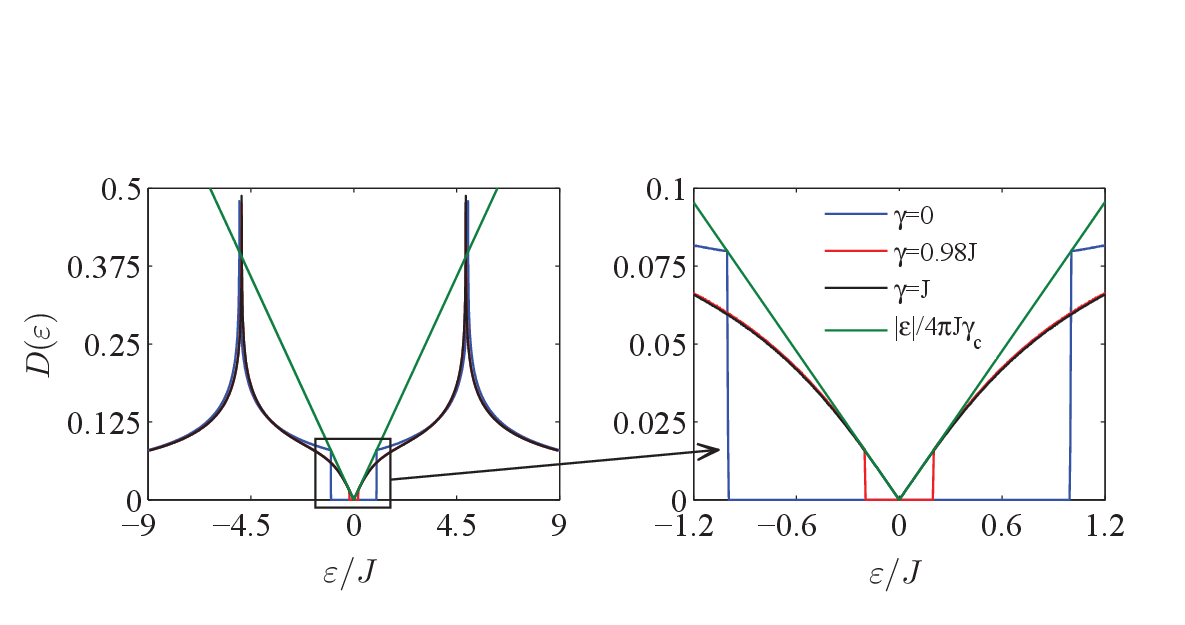}
\caption{(color online). DOS per unit cell as a function of energy (in units
of $J$) computed from the energy dispersion Eq. (\protect\ref{E2D}) with
several typical values of $\protect\gamma =0$ (blue), $0.98J$ (red), $J$
(black). And here we set $T=5J$. Also shown is a zoom-in of the densities of
states close to the zero-energy point, which can be approximated by $D\left(
\protect\varepsilon \right) \propto \left\vert \protect\varepsilon %
\right\vert $. The approximate expression in Eq. (\protect\ref{app DoS}) is
also plotted (solid green) as comparison.}
\label{fig4}
\end{figure}

We introduce density of states (DOS) to characterize the band structure. DOS
is essentially the number of different states at a particular energy level
that electrons are allowed to occupy, i.e., the number of electron states
per unit volume per unit energy. DOS calculations allow one to capture
various electronic properties, such as specific heat, paramagnetic
susceptibility, and other transport phenomena of conductive solids. The DOS $%
D\left( \varepsilon \right) $ of energy bands for a square lattice can be
expressed as follows
\begin{equation}
D\left( \varepsilon \right) =\frac{1}{4\pi ^{2}}\int \!\!\!\int_{\mathrm{B}%
}\delta \left[ \varepsilon -\varepsilon ^{\pm }(k_{x},k_{y})\right] \mathrm{d%
}k_{x}\mathrm{d}k_{y},
\end{equation}%
which describes the number of states per unit energy per unit cell and
therefore the function is properly normalized to $\int_{\mathrm{B}}D\left(
\varepsilon \right) \mathrm{d}\varepsilon =2$. Due to the symmetry of
spectrum, we have $D\left( \varepsilon \right) =D\left( -\varepsilon \right)
$. Here the densities of states for different $\gamma $ are plotted in Fig. %
\ref{fig4}. In the vicinity of $k_{\mathrm{c}}$ and considering the case $%
0\leqslant \gamma _{\mathrm{c}}-\gamma \ll J$, Eq. (\ref{hyperboloid})
allows us to derive an approximate expression for the density of states
\begin{equation}
D\left( \varepsilon \right) =\left\{
\begin{array}{cc}
\frac{1}{4\pi J\gamma _{\mathrm{c}}}\left\vert \varepsilon \right\vert , &
\left\vert \varepsilon \right\vert \geqslant \sqrt{\gamma _{\mathrm{c}%
}^{2}-\gamma ^{2}} \\
0, & \left\vert \varepsilon \right\vert <\sqrt{\gamma _{\mathrm{c}%
}^{2}-\gamma ^{2}}%
\end{array}%
\right. ,  \label{app DoS}
\end{equation}%
which is a linear function of energy. We plot this expression in Fig. \ref%
{fig4} as comparison. It indicates that $D\left( \varepsilon \right) $\
shows a semimetallic behavior as that in graphene.

\section{Summary}

\label{Summary}In conclusion, we have studied the connection between an
arbitrary Hermitian tight-binding model with $\mathcal{C}$ symmetry and its
non-Hermitian counterpart with $\mathcal{CT}$ symmetry. It has been shown
that such a kind of non-Hermitian Hamiltonian is pseudo-Hermitian, providing
a way to generate a class of non-Hermitian models with a tunable full real
band structure by adding additional imaginary potentials. Based on the exact
results, it is found that, the eigenvalues and eigenvectors of the
non-Hermitian Hamiltonian can be easily obtained from those of its parent
Hermitian Hamiltonian. The reality of the spectrum is robust to the disorder
due to the protection of energy gap. Furthermore, as an illustrative
example, we investigate the band structure of a two-layer square lattice
model with staggered imaginary potentials. We find that a tunable cone band
structure can be achieved. It should have wide applications in non-Hermitian
synthetic graphene-like materials.

\acknowledgments We acknowledge the support of the National Basic Research
Program (973 Program) of China under Grant No. 2012CB921900 and CNSF (Grant
No. 11374163).\newline
\newline


\end{document}